\documentclass[a4paper,fleqn,usenatbib]{mnras}

\usepackage{newtxtext,newtxmath}

\usepackage[T1]{fontenc}
\usepackage{ae,aecompl}

\usepackage{graphicx}	
\usepackage{amsmath}	
\usepackage{amssymb}	
\usepackage{xcolor}

\title[H$_2$--bearing GRB-DLAs]{Physical conditions in two high-redshift H$_2$-bearing GRB-DLAs, 120815A and 121024A.}

\author[Shaw \& Ferland]{
Gargi Shaw$^{1}$\thanks{E-mail: gargishaw@gmail.com},
{G. J. Ferland$^{2}$}
\\
$^{1}$Department of Astronomy and Astrophysics, Tata Institute of fundamental research,\\
Homi Bhabha Road, Navy Nagar, Colaba, 
Mumbai 4000005, India\\
$^{2}$Department of Physics and Astronomy,
University of Kentucky,Lexington, KY 40506, USA\\}

\date{Accepted ... . Received ; in original form }

\pubyear{2020}

\begin{document}
\label{firstpage}
\pagerange{\pageref{firstpage}--\pageref{lastpage}}
\maketitle

\begin{abstract}
The gamma-ray burst (GRB) afterglows provide an unique opportunity to
study the interstellar medium (ISM) of star-forming galaxies at high-redshift. 
The GRB-DLAs (damped Lyman-$\alpha$ absorbers) contain a large neutral hydrogen column density, N(H I), and are observed
against the GRB afterglow. A large fraction of GRB-DLAs show presence of molecular hydrogen (H$_2$) which is an indicator of star-formation. 
Hence it is important to study those GRB-DLAs which have H$_2$ lines 
to decipher and understand their physical conditions.  The GRB-DLAs 121024A and
120815A, situated at redshift 2.30 and 2.36 respectively, are two
such important H$_2$-bearing GRB-DLAs. Besides H$_2$, these
two GRB-DLAs also show many metal lines. In this work we have carried out a detail numerical study on the H$_2$ lines, as well as on those metal lines, 
in GRB-DLAs 121024A and 120815A self-consistently. We use the spectral
synthesis code CLOUDY for this study. This modeling helps us to determine the 
underlying physical conditions which give rise such lines and hence to understand these two GRB-DLAs in much more detail than any other previous investigation. 
We find that the hydrogen densities for these two H$_2$-bearing DLAs are $\geq$ 60 cm$^{-3}$. Moreover our study infers 
that the linear sizes are $\leq 17.7$ pc for these two 
GRB-DLAs, and the mean gas temperatures averaged over the cloud thickness, are $\leq$ 140 K.
Overall, we find  that these two H$_2$-bearing GRB-DLAs are denser, cooler and smaller compared to those  without H$_2$.
\end{abstract}


\vspace*{0.05in}



\section{Introduction}

Molecular hydrogen (H$_2$) is the first neutral molecule to be formed in the universe, and it is the most abundant and the main
constituent of molecular clouds where star formation takes place \citep{{1992Burton},{2005Tielens},{Bigiel2008}}. 
Furthermore, H$_2$ controls most of the chemistry 
in ISM through its ionic or neutral form. In addition to these, level populations of H${_2}$
in various levels can be used as tracers for physical conditions. For example, in the well-shielded H$_2$ gas
the lower rotational levels of the ground state of H$_2$ are mostly 
collisionally dominated, 
hence they are in LTE, and can 
be used to infer gas temperature \citep{1992Abgrall}. Whereas, the higher levels are generally populated by non-thermal processes and can be used to estimate ambient radiation 
field \citep{{1996Draine},{2005Shaw}} and the cosmic ray ionisation rate of hydrogen \citep{1997Tine}. As a result, H$_2$ spectra 
provide an excellent opportunity to probe star formation and chemical 
enrichment of galaxies ranging from local to high-redshift (z). 

In Milky Way, a
large neutral hydrogen column density N(H I) is in general associated
with H$_2$ \citep{{1977Savage},{2017Winkel},{2017Marasco}}. If the physical conditions of high-z galaxies are similar to that of Milky Way, it is expected
that large N(H I) regions at high-z galaxies are also
associated with H$_2$. Therefore, chemical evolutions of these high-z galaxies can also be studied through systems with large N(H I).
The damped Lyman-$\alpha$ systems (DLAs) are
such systems with a large N(H I) \citep{2005Wolfe} and which are
also the main reserviour of neutral gas at high-z. 

DLAs can be probed using
absorption spectroscopy against distant bright sources, such as
quasi-stellar objects (QSOs) or gamma-ray bursts (GRBs).
DLAs observed against the QSOs
are called QSO-DLAs, whereas, DLAs detected against the GRB afterglow
are called GRB-DLAs. The long GRBs, with a duration greater than two seconds,
are believed to be originated from the core-collapsed supernova.
Since massive stars are short lived and are located in star-forming
regions of galaxies, it is apparent that GRBs are also associated with
star-forming regions of a galaxy \citep{1993Woosley}. An excellent review of GRB-DLAs in the Swift era is provided by \citet{2017Schady}. 

To date, many QSO-DLAs and GRB-DLAs have been observed by several
groups \citep{{Prochaska2001}, {Srianand2005b}, {2003Ledoux}, {Daniel2008}, {2016Toy}, {2009Ledoux}, {2019Heintz}, {2019Bolmer}} etc. The GRB-DLAs are
smaller in number than QSO-DLAs as GRBs are transitory in
nature. However, the GRB-DLAs show larger N (H I) and higher
metallicity than those of QSO-DLAs \citep{2007Prochaska}. Here, metallicity relative to solar is expressed as 
[X/H]=log[N(X)/N(H)]-log[N(X)/N(H)]$_{\odot}$ (with X= Zn, or Si, or S).
It was also
noticed by \citet{2015Cucchiara} that at high-z the average
metallicity of QSO-DLAs decline at a faster rate than GRB-DLAs.
In general, QSO-DLAs probe diffuse gas and hence they are not suitable to
study star formation. On the contrary, since GRBs are located inside the galaxies, GRB-DLAs probe inner regions of galaxies \citep
{2008Fynbo}, and consequently, GRB-DLAs are much suitable for studying star-forming regions. However, recent studies of extremely saturated DLAs 
(ESDLA) \citep{{2014Noterdaeme}, {2015aNoterdaeme}, 
{Balashev2017}, {Ranjan2018}, {2019Bolmer}, {Ranjan2020}}
indicate that ESDLAs and GRB-DLAs likely represent the same population of the galaxies, probed by small impact parameters, and hence can probe 
star-forming regions. Furthermore, recently, a 
huge data is also available on GRB-DLA host galaxies up to a very 
high redshift (z$\approx$ 6) \citep{{2015Hartoog},{2019Bolmer}}.

Initially, \citet{2007Tumlinson} studied five GRB-DLAs but did not find H$_2$ in spite of large N(H I). They concluded that this 
lack of H$_2$ may be due to a combination of low metallicity and an FUV radiation field of 10-100 times the Galactic mean field. Later, \citet{2009Ledoux} 
observed seven z > 1.8 GRB afterglows with VLT/UVES but they also did not find H$_2$ in their sample. They explained the lack of detected H${_2}$ 
through the low metallicities, low depletion factors, and moderate particle densities of these 
systems. Though \citet{2007Tumlinson} and \citet{2009Ledoux} differed regarding the FUV radiation field, they both agreed on the low metallicity of  
these systems. \citet{2009Ledoux} estimated a particle density of 5-15 cm$^{-3}$, a linear cloud size of 520$^{+240}_{-190}$ pc, and the kinetic temperature 
>1000 K for these 
seven z > 1.8 GRB afterglows of their sample which did not show H$_2$. Later \citet{2016Toy} observed 
a sample of 
45 GRB-DLAs in the redshift range of 2 to 6 and they found that DLA counterpart star formation rates (SFRs) are not correlated with either redshift or H I 
column density. 

However, some other GRB-DLAs, which harbour H$_2$, were found, namely, GRB-DLA 80607, GRB-DLA 121024A, GRB-DLA 120815A and GRB-DLA 120327A at 
redshifts 3.03, 2.30, 2.36 and 2.8 \citep{{2009Sheffer}, {2015Friis}, {2013Kruhler}, {2014DElia}}.
Recently \citet{2019Bolmer} have observed 22 GRB-DLAs with VLT/X-shooter for z > 2. In their sample, they found H$_2$ absorption lines in 6 out of these 22 
GRB-DLAs, 
which also include GRB-DLA 121024A, GRB-DLA 120815A and GRB-DLA 120327A. They concluded that there is no lack of detected H${_2}$ for GRB-DLAs and the detection rate 
is much higher in GRB-DLAs than the QSO-DLAs. It has been noted that for GRB-DLAs 
with log N (H I) > 21.7, the detection of molecular hydrogen increases \citep{2019Bolmer}. The same increase in the detection 
rate at high HI is also seen in QSO-DLAs \citep{{2015aNoterdaeme},{2018Balashev}}. Earlier, such conversion of HI to H$_2$ was analytically studied 
\citep{{1988Sternberg},{2014Sternberg}}, and observationally constrained for both low and high 
redshifts \citep{{1977Savage}, {2016Welty}, {2015aNoterdaeme}, {2018Balashev}}.
It is thus natural to ask whether the physical conditions of 
H$_2$-bearing DLAs are quite different than those without H$_2$, and if so, how to probe and then understand that. One of the plausible 
answers for detectable H$_2$ could be 
the presence of dust grains, higher metallicity, and higher density. Dust grains are very important as their surfaces act as a 
catalyst for efficient H$_2$ formation. As an example, it has 
been suggested by many observers \citep{{Srianand2005a},{Srianand2005b}} that 
the H$_2$-bearing QSO-DLAs might have higher density and higher dust content compared to the non H$_2$-bearing QSO-DLAs.  

A detail numerical spectroscopic modeling of GRB-DLAs considering all the possible microphysics of H$_{2}$ is thus quite crucial to determine 
the underlying physical 
conditions which give rise such lines, and hence to understand these systems which harbour H$_{2}$.
Previously, \citet{2008Whalen} had carried out a numerical simulation to understand the absence of H$_2$ in GRB-DLAs. However, so far, no one has performed any  
detail microscopic modelling of such systems which harbour H$_2$. Previously, we successfully carried out several detail modelings of 
QSO-DLAs \citep{Srianand2005a, Srianand2005b, 2016Shaw, 2018Rawlins} with detected H$_2$ lines. Following that in this work, 
we employ the spectral synthesis code CLOUDY \citep{Ferland2017} for a microscopic detail modelling with an aim to understand the physical 
conditions of H$_2$-bearing GRB-DLAs. 

Our calculation incorporates 
detail microphysics both at atomic and molecular levels including collisional physics and line shape theory. 
In this work we choose following two systems: GRB-DLAs 121024A and 120815A, and study them self-consistently.  
We select these two GRB-DLAs as they have higher molecular fraction with log(f)$\approx$ $-1.14 \pm$ 0.15 and $-1.4$, 
respectively, and so will serve our purpose best to distinguish H$_2$-bearing  from non-H$_2$-bearing GRB-DLAs. 
In addition to H$_{2}$, these two DLAs 
show numerous metal 
lines together with the rest frame UV absorption lines of H$_2$. 
Along with H$_2$ lines, above mentioned extra lines are also modeled self-consistently in our calculation. 

This article is organized as below: in section 2 we give details of our calculation using CLOUDY. Detail of microphysics incorporated in our 
calculation is also provided. In section 3 we present the findings from this study, first providing results for GRB-DLA 120815A and then for 
GRB-DLA 121024A. Summary and conclusions are presented in section 4.

\section{Calculations}

CLOUDY is a self-consistent stationary 
micro-physics code based on $\textit {ab initio}$ calculation of thermal, ionisation, and chemical balance of non-equilibrium gas and dust 
exposed to a source of radiation. It predicts column densities of various atomic and molecular species 
and resultant spectra covering the whole range of EM radiation and 
vice verse using a  minimum number of input parameters.  It has a state-of-the art detailed H$_2$ network \citep{2005Shaw} embedded which 
is very helpful for modelling  environments with H$_2$. The H$_2$ network includes 301 rovibrational levels within the ground electronic state and also 
the rovibrational levels within the lowest six electronic excited states. 

Here, we briefly discuss the main formation and destruction processes
of H$_2$ that are implemented inside CLOUDY and affect the H$_2$
level populations. Since the formation process of H$_2$ is exothermic and
the resulting energy for such process is close to its dissociation energy, H$_2$ formation does not simply take place in gas phase by interaction of 
two H atoms.
It requires a mechanism to take away some of the formation energy to form a stable H$_2$ molecule. In a dusty environment, dust plays the role to 
share that formation energy. As a result, H$_2$ is mainly
formed on dust grain surfaces where dusts can act as a catalyst and take
away some of the formation energy. In a dust-free environment, H$_2$ can still
form through an exchange reaction of H with H${^-}$ \citep{1991Launay} and
H${_2}{^+}$ \citep{2002Kristic} but the reaction rates are smaller. The main destruction and excitation mechanism for H$_2$
is the photoexcitation. In this process, H$_2$ absorbs Lyman
and Werner band photons and gets excited to higher electronic
levels. Of these excited populations, 10-15\% comes down to the
continuum of ground electronic state and get dissociated. The rest
populates higher vibrational levels of the ground electronic state
\citep{1992Abgrall}.  Besides this, the formation process is exothermic, hence the dust
grains also play a crucial role in the distribution of level population of
newly formed H$_2$ on dust grain surfaces \citep{{1987Black},
  {1995LeBourlot}, {1996Draine}, {2001Takahashi}}. Note that deep into the cloud, excitation by secondary electrons produced by 
  cosmic ray is also an important
H$_2$ level excitation mechanism \citep{1997Tine}.  
All these above mentioned processes are incorporated inside CLOUDY and details are given in \citet{2005Shaw} and \citet{2012Gay}. 
It is well known that depending on the nuclear spin orientations, H$_2$ can be in one of the two quantum states, namely, ortho or para. 
In an ortho state, the nuclear spins are aligned parallel, whereas in a para state, they are anti-parallel. 
Note that radiative transitions are not possible between ortho and para states of the ground electronic level and only reactive 
collisions with H$^{0}$, H$^{+}$, and H$^{+3}$ \citep{1994Sun, 1990Gerlich, 1991Bourlot} are 
capable to make transitions between these quantum states. In our H$_2$ network, CLOUDY includes both reactive and non-reactive collisions. 
For details, see \citep{2005Shaw}. 

\subsection{Modelling}
\label{Modelling}For all the models that we incorporated in this work, we assume a plane-parallel constant pressure gas exposed to radiation. 
We vary a few free parameters, 
as will be discussed below, to match the predicted column densities with the observed column densities of various atoms, ions and state specific H$_2$. 
The chosen radiation field has three components: a meta-galactic radiation at appropriate redshift  \citep{2012Haardt}, a synchrotron radiation 
(power-law continuum) arising from GRB afterglow and a blackbody radiation from $\textit {in situ}$ star formation. 
Our GRB-DLA models are similar to our previous QSO-DLA models \citep{{Srianand2005a}, {2016Shaw}, {2018Rawlins}} except that here we have 
an additional synchrotron 
radiation arising from GRB afterglows. While the gas is exposed to UV radiation from everywhere it is exposed to GRB afterglow only from one side. 

In the current CLOUDY set up, it is not possible to consider both two-sided and one-sided radiation fields simultaneously in a same model.
    Hence we consider following two cases separately: Case 1, where we assume that the gas cloud is irradiated from both sides; and 
    Case 2: where we assume that the gas cloud is irradiated from one side. We present our results for both the cases. 
    The models presented here show that this does not effect the conclusions. However, it need to be noted that the geometry of 
    the cloud is 
    unknown and might be unknowable. The time history of the cloud is not known and we are also using a stationary code. These can also affect the results.

\subsection{Input parameters}
\label{Input parameters}
In this work, we use the dimensionless ratio of hydrogen-ionising photon $Q(H)$ (s$^{-1}$) to total-hydrogen
densities $n(H)$ (cm$^{-3}$) in order to quantify the radiation field, and denote it by ionisation parameter, $\chi$, as, 
\begin {equation}
\chi = \frac{Q(H)} {4\pi {r_0}^2 c\, n(H)} . 
\end {equation}
Here $r_0$ and $c$ are the separation (cm) between the center of the source
of ionizing radiation and the illuminated face of the cloud, and the speed of light, respectively. 

We consider both graphite and silicate grains in the 
calculations with MRN \citep{1977Mathis} size distribution over ten size bins in the range 0.005 to 0.25 micron (ISM grains). 
For MRN grains, the grain size distribution 
varies as $a^{-3.5}$, where $a$ is the 
radius of the grain. 
Beside that, dusts provide heating through photoelectric heating, as well as it shields interior regions of molecular cloud from 
far-ultraviolet (FUV) radiation. The detail grain physics included in CLOUDY was described in \citet{2004vanhoof}.  It has been observed that dust 
is correlated with metallicity,  and in general, metallicity is anti-correlated with redshift. It is believed that Fe is strongly depleted 
on dust grains whereas Zn is not depleted. Hence, one can estimate the metallicity and dust-to-gas ratio with the help of Zn/H and Fe/Zn ratios 
\citep{Prochaska2002}. In our models, both metallicity and dust-to-gas ratio are derived using  column densities of Zn, Fe and equation no. 1 from 
\citet{Prochaska2002}.

As mentioned earlier, cosmic ray affects the level populations of H$_2$ via secondary electrons which are produced by 
cosmic ray ionisation \citep{1997Tine}. 
However, the cosmic ray ionisation rate is not the same 
along all line-of-sights
\citep{{2007Indriolo},{2008Shawa}}. Since not much is known on the cosmic ray ionisation rate at high-redshift, 
we consider this rate as a free parameter in our models. 

We also treat the total hydrogen density, which includes all forms of hydrogen bearing chemical species, as a free parameter in our models. To be noted that CLOUDY 
requires at least one suitable stopping criterion for each model and it will stop calculations at some depth of the cloud depending on the stopping criterion.
These stopping criteria vary depending  on particular observed quantities that one match with.
For all our case 2 models, we consider the observed total H$_2$ column densities as 
the stopping criterion, i.e., CLOUDY will continue its calculation till a depth of the cloud where the predicted total H$_2$ column density 
matches well with the corresponding observed value. Whereas, for case 1, we stop the model at a depth in the cloud where the total H$_2$ 
column density equals half of the observed H$_2$ column density. Finally we multiply our predictions by 2 to mimic the situation where the cloud is 
irradiated from both sides.

A built-in optimization program, based on phymir algorithm \citep{vanhoof1997}, which calculates a 
non-standard goodness-of-fit estimator $\chi^2$ and minimises it by varying input parameters, is employed to identify the best model.
Sometimes, to reach to the final model of a calculation, a few parameters are fine tuned so that the observed data can be matched maximally.
This optimisation program is user friendly and has been utilised extensively in many previous works 
\citep{{Ferland2013}, {2000Srianand}, {2006Shaw}, {2018Rawlins}, {2019Shaw}}.

    In this work, using various parameters as discussed above, our main focus is to model the observed state-specific H$_2$ column densities self-consistently 
together with other observed column densities of ionic, atomic and molecular species to understand the underlying physical conditions 
which play pivotal roles to
generate them. Below we discuss our findings.
  
\section{Results}
In this section, results and main findings are elaborated. First GRB-DLA 120815A is discussed in detail, followed by GRB-DLA 121024A.
\subsection{GRB-DLA 120815A}
\label{subsec:GRB-DLA 120815AA}
For the GRB-DLA 120815A, H$_2$ lines were first detected at z = 2.36 by \citet{2013Kruhler}. They observed the line-of-sight using X-shooter at VLT. 
H$_2$ absorption lines were observed in the rest frame Lyman and Werner bands (11.2--13.6 eV) which have been shifted to optical band due to redshift and 
 hence could be easily 
observed by X-shooter which is an optical telescope.
They reported H$_2$ column densities for 
 the first four rotational levels of ground state. The neutral carbon and H$_2$ gets photo-dissociated by photons of the same energies, and hence
 the neutral carbon generally co-exists with H$_2$ \citep{{2007Noterdaeme},{2018Noterdaeme},{2009Jorgenson}}. As expected, the neutral carbon was also observed in this
 GRB-DLA.
 Beside H$_2$, they also detected various metal lines, such as Zn, S, Si, Mn, Fe, Ni in their first ionisation states.
 The observed metallicity ($[$Zn/H$]$) and dust depletion ($[$Zn/Fe$]$) were found to be $-1.15 \pm 0.15$ and $1.01 \pm 0.10$, respectively. Using these the 
 dust-to-gas ratio is calculated to be $\approx$ 
0.06 of the local ISM value, and is used in our models. They also found the broadening 
 parameter of the H$_2$ lines to be $8.7 \pm 0.6$ km s$^{-1}$. In our model, we fix the microturbulence/broadening parameter to the observed value 8 km s$^{-1}$.

\begin{figure}
	
\vspace*{-1.1in}
\hspace*{-0.25in}
\includegraphics[scale=0.5]{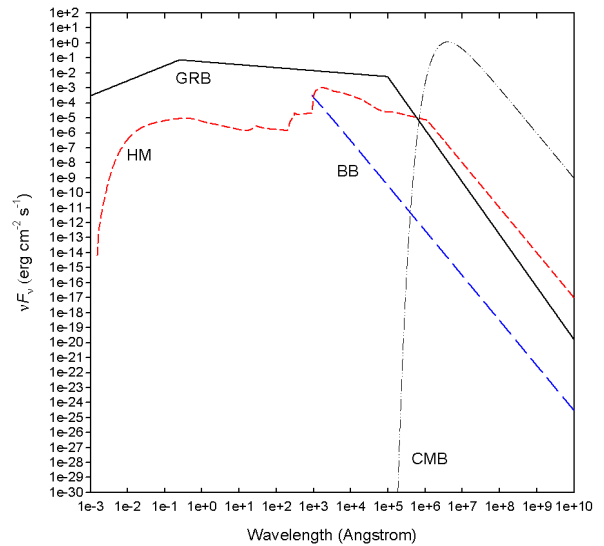}
\vspace*{-1.5in}
\caption{The SED of the individual components of incident radiation field for GRB-DLA 120815A. The dash-dotted (black) and medium-dashed 
(red) lines represent
CMB and  Haardt-Madau metagalactic radiation field at z=2.36, respectively. The long-dashed (blue) line represents extinguished (no H-ionising radiation) 
stellar 
radiation generated using black body radiation at 10$^{5}$ K. The solid black line represents the powerlaw continuum from the GRB. 
The powerlaw continuum behaves 
as $\nu$$^{5/2}$ at lower energy to account for self-absorbed synchroton \citep{1979Rybicki}, $\nu$$^{-2}$ at 
higher energies and -0.8 between 10 micron and 50 KeV.}
\label{fig:fig_conti}
\end{figure}

As mentioned previously, in our model we consider that radiation consists of a meta-galactic background at z=2.36, a blackbody radiation 
from an $\textit in situ$  star formation and a power law continuum due to afterglow. \citet{2013Kruhler} found a power 
law slope of $-0.78$ as the best-fit value. Furthermore, recently, 
\citet{2018Li} have published 
a large catalogue of multi-wavelength GRB afterglows with spectral indices of 70 GRBs, and they have mentioned -0.78$\pm$0.01 as the spectral index for GRBs 
120815A. Based on these facts, we choose $-0.8$ as the spectral index for GRB. As an example, in Fig.\ref{fig:fig_conti} we plot all the components of the 
SED for 
GRB-DLA 120815A (case 1). The corresponding UV field is equivalent to 7 Habing. Case 2 also has the similar features with corresponding UV field 6.6 habing.
In Table \ref{tab:table 1} we list the model parameters corresponding to our best model and the second and third columns represent Case 1 and 
      Case 2, respectively. In case 1, our model  predicts a hydrogen density of 440 cm$^{-3}$. Whereas, for case 2 the predicted 
      hydrogen density is lower, 230 cm$^{-3}$.
For other elements, whose lines were observed, we vary their abundances in our model to match their observed values while keeping the abundances
of other elements at 0.07 of the ISM value. Final values of metal abundances of Zn, S, Si, Fe, Ni, Cr and Mn, that we obtain from our best fit model, 
match well with the observed data (within the observed error bars) by  
\citet{2013Kruhler}. In addition to these,
we also predict the abundances of C, O, N and Mg. \citet{2013Kruhler} have measured the A$_v$ = 0.15 $\pm$ 0.02 along this sight line. 
However, we predict  A$_v$ = 0.32 and 0.31 for case 1 and case 2, respectively. These are higher than the observed value.
We also find that the cosmic ray ionisation rate of hydrogen along this line-of-sight is  approximately 2 $\times$ 10$^{-16}$ s$^{-1}$, similar to the average galactic 
value reported by \citet{2007Indriolo}.  In Table \ref{tab:table 2} we compare the observed data and our model 
predicted column densities in two different columns. To be noted that in addition to the observed species, observable amount of OH and OH$^+$ 
with column density > 10$^{13}$ cm$^{-2}$ are also predicted by our models.
One can see that our model predicts the 
rotationally resolved H$_2$ column densities within the observed range, except for rotation level $\textit J$ = 2.

\begin{table}
	\centering
	\caption{Physical parameters for GRB-DLA 120815A using CLOUDY.}
	\label{tab:table 1}
	\begin{tabular}{lcr} 
		\hline
		Physical parameters & best values & best values\\
		&Case 1 & Case 2\\
		\hline
		Power law: log($\chi$) & -4.2 & -4.0\\
		Black body: Temperature (log K), log($\chi$) & 5.0, -1.8 & 5.3, -2.6\\
		Density n(H) ($cm^{-3}$) & 440 & 230\\
		Cosmic ray ionisation rate (10$^{-16}$ s$^{-1}$) & 1.9 & 1.9 \\
		$[$Fe/H$]$  & -2.24 & -2.27\\
		$[$Mg/H$]$ & -1.93 & -1.89\\
		$[$C/H$]$ & -2.39 & -2.39\\
		$[$Zn/H$]$& -1.28 & -1.27\\
		$[$Ni/H$]$ & -2.17 & -2.19\\
		$[$Mn/H$]$ & -2.27  &-2.27      \\
		$[$Cr/H$]$ &  -2.07 & -2.04  \\
		$[$S/H$]$ & -1.48 & -1.20\\
		$[$N/H$]$& -1.18 & -1.18\\
		$[$O/H$]$& -1.33 & -1.33\\
		\hline
		\end{tabular}
\end{table}
		
\begin{table}
	\centering
	\caption{Comparison of observed and predicted column densities (in log scale) for GRB-DLA 120815A using CLOUDY.}
	\label{tab:table 2}
	\begin{tabular}{lccr} 
		\hline
		Species  & observed  & predicted & predicted \\
		&(cm$^{-2}$)&(cm$^{-2}$)&(cm$^{-2}$)\\
		& & Case 1 & Case 2\\
		\hline
		H I & $21.95\pm 0.10$& 22.10 &22.08 \\
		H$_2$ & $20.54\pm 0.13$& 20.54 & 20.54\\
		Mg I & $13.54\pm 0.05 $& 13.48 & 13.49\\
		Ni II & $14.19 \pm 0.05$ & 14.17 & 14.17\\
		Zn II & $13.47 \pm 0.06 $& 13.44 & 13.44 \\
		C I & $13.41 \pm 0.11 $& 13.41 & 13.43\\
		Fe II & $15.29 \pm 0.05$ & 15.34 & 15.26\\
		Mn II & $13.26 \pm 0.05 $& 13.22 & 13.16\\
		Cr II & $13.75 \pm 0.06 $& 13.73 & 13.74 \\
		S II & $\le 16.22\pm 0.25$ & 15.90 & 16.17\\
		Si II &$\ge 16.34\pm 0.16 $& 16.00 & 15.85\\
		CO & < 15.0 & 11.60 & 11.50\\
		OH & --& 13.85 & 13.85\\
		OH$^+$&--& 13.16 & 13.36\\
		HCl & --& 12.67 & 12.46\\
		H$_2$O &-- &12.54 & 12.42\\
		H$_2$O$^+$ &--&12.41 & 12.46\\
		H$_3$$^+$ & -- &12.42 & 12.51\\
		H$_2$(0) &19.84 $\pm$ 0.33  & 20.02 & 19.98\\
		H$_2$(1) & 20.43 $\pm$ 0.12 & 20.37 & 20.38\\
		H$_2$(2) & 16.76 $\pm$ 0.50 & 18.89 & 19.01\\
		H$_2$(3) & $\le$ 19.01 & 17.91 & 17.99\\
		H$_2$(total) &20.54 $\pm$ 0.13 & 20.54 &20.54\\
		\hline			
	\end{tabular}
\end{table}

Different parameters affect the predicted column densities to different amounts. To see the variation in predicted column densities due 
to change in parameters, we vary the important parameters by 0.3 dex around their values used in the best model (Case 1). In Table 3 we present this 
change in predicted column densities (in dex). We would like to emphasis here that the stopping criterion, i,e., total H$_{2}$ column density, remains the same for 
each set of parameters. Our calculation shows that among all parameters the effect of  
dust grains is the maximum. Similar trend is observed for case 2.

\begin{table}
	\centering
	\caption{Variation in column densities of GRB-DLA 120815A (in dex) for variation of different input parameters (Case 1).}
	\label{tab: table 3}
	\begin{tabular}{lccr} 
		\hline
		Species  & $\Delta$n(H) & $\Delta$grain  & $\Delta$log($\chi$)\\
		 &&&power law\\
		 & 0.3 dex& 0.3 dex  & -0.3 dex\\
		\hline
		H I &0.02 &-0.-30 & -0.20\\
		Mg I &-0.01 & -0.35&0.04\\
		Ni II & -0.07&-0.38 & -0.22\\
		Zn II &0.01 &-0.28 & -0.20\\
		C I & -0.02&-0.32& -0.04\\
		Fe II & 0.01 &-0.28 &-0.22\\
		Mn II &0 & -0.27&-0.22\\
		Cr II &0.01 &-0.28 &-0.21 \\
		S II &0.02 & -0.28&-0.22\\
		CO &0.10  &0.21 &0.27\\
		OH & 0.08&0.22 &0.05\\
		OH$^+$&0.01 &-0.10 &0.25\\
		HCl &0.03&-0.06&0.03\\
		H$_2$O & 0.14& 0.39&0.21\\
		H$_2$O$^+$ &0.85 & 0.14&-0.13\\
		H$_3$$^+$ &0  &0.18 &0\\
		H$_2$(0) &-0.03 & -0.05&0.07\\
		H$_2$(1) & 0& 0.01&-0.03\\
		H$_2$(2) & 0.10& 0.16&-0.20\\
		H$_2$(3) & 0.24 & 0.13&-0.31\\
		\hline		
	\end{tabular}
\end{table}

\begin{figure}
	
  \vspace*{-1.1in}
\hspace*{-0.3in}
\includegraphics[scale=0.4]{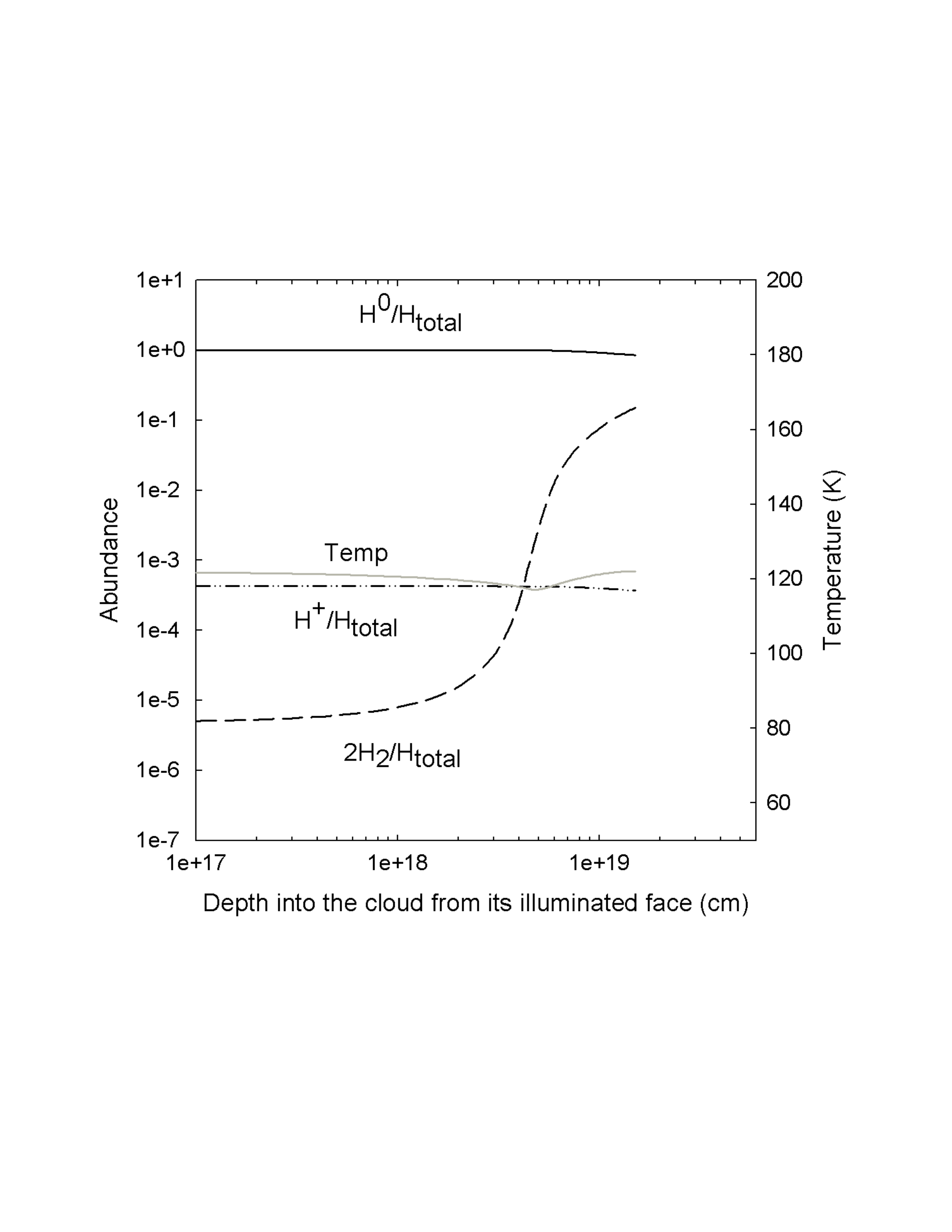}
\vspace*{-1.2in}
\caption{Temperature and H$^0$/H$_{total}$, H$^+$/H$_{total}$, 2H$_2$/H$_{total}$ are plotted as a function of distance into GRB-DLA 120815A (Case 1). 
The temperature axis is shown in the right. This plot shows
    half of the cloud. The other half is just the reflection of this portion.}
    \label{fig:fig_ad1}
\end{figure}

\begin{figure}
\vspace*{-0.5in}
\hspace*{-0.3in}
\includegraphics[scale=0.4]{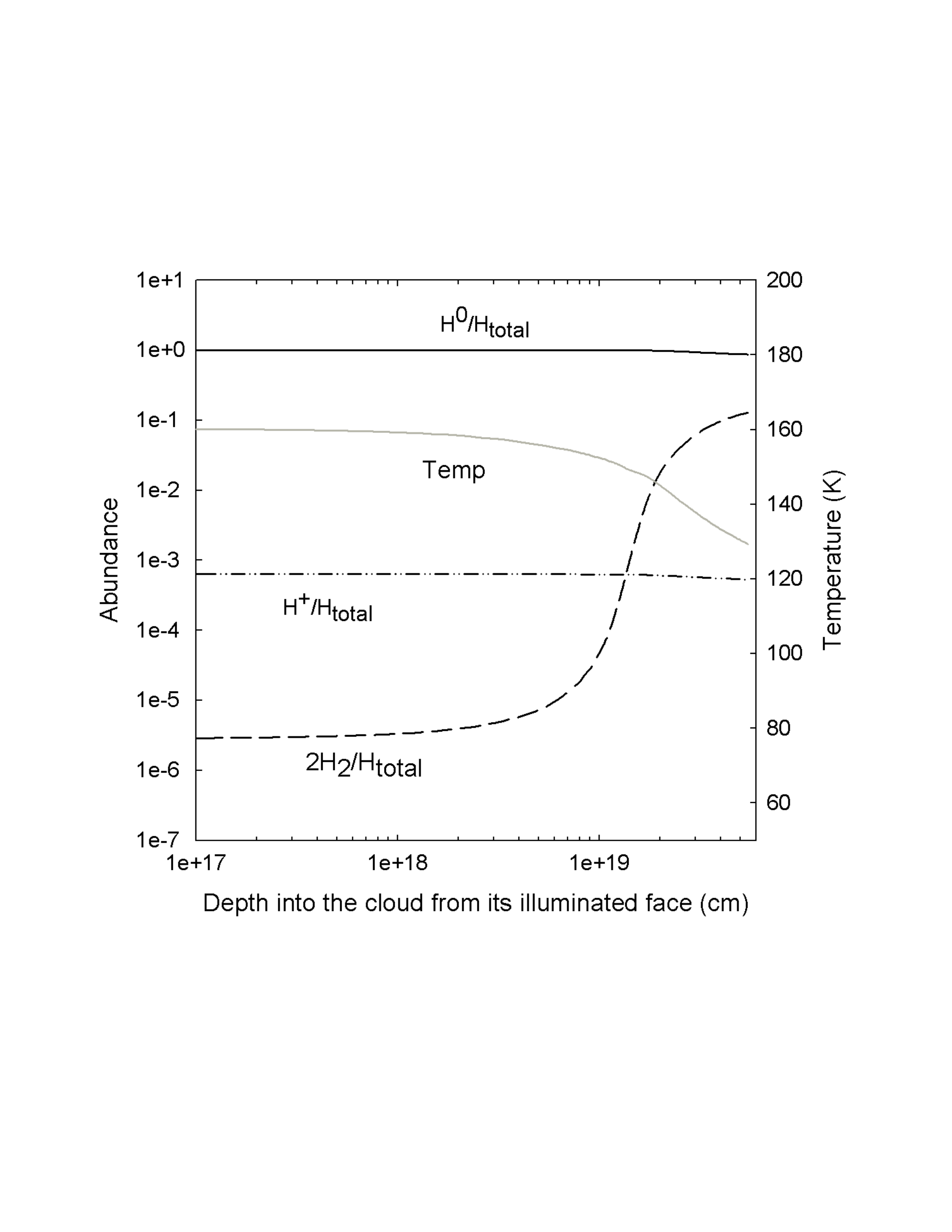}
\vspace*{-1.2in}
    \caption{Temperature and H$^0$/H$_{total}$, H$^+$/H$_{total}$, 2H$_2$/H$_{total}$ are plotted as a function of distance into GRB-DLA 120815A (Case 2). 
    The temperature axis
    is shown in the right.}
    \label{fig:fig_as1}
\end{figure}

				
As we know that numerical models can be utilized to predict various physical
conditions across the cloud which are not possible by todays' observations.
In our model we also calculate the abundances of 
H$^0$/H$_{total}$, H$^+$/H$_{total}$, 2H$_2$/H$_{total}$ and the variation of gas temperature across the cloud, Figs. \ref{fig:fig_ad1} 
and \ref{fig:fig_as1} show those 
for Case 1 and case 2, respectively. Fig.1 shows half of the cloud, and  
the other half is its reflection.
Our best models predict the size of the GRB-DLA 120815A to be 9.8 and 17.7 parsec for case 1 and case 2, respectively.
Interestingly, this linear size is much smaller 
than that reported by \citep{2009Ledoux} for GRB-LAs with no H$_2$. 
The gas temperature is one of the important physical conditions for any astrophysical environment. 
It is well known that in the warm phase of the ISM,  T$>$1000 K and n$<$10 cm$^{-3}$. Whereas, in the dense cold phase T$\sim$100K and n$>$100 cm$^{-3}$. 
For Case 1, the predicted gas 
temperature remains almost constant  
near 120 K. On the contrary, for case 2 the  gas temperature ranges from 157 K to 129 K with an average of 142 K over the linear extension of the cloud.
 In a collisionally dominated environment, $\textit J$ = 1 and 0 levels of H$_2$ 
are generally thermalised and can be used to determine the surrounding gas temperature. This excitation temperature, T$_{10}$, is calculated using the 
following  equation,
 \begin {equation}
T_{10} = -170.5 \left[ln \frac{N(J = 1)} {9N(J = 0)}\right]^{-1} K.    
\end {equation}
Here ${N(J = 1)}$ and ${N(J = 0)}$ are column densities in the $\textit J$ = 1 and 0 levels of H$_2$, respectively, 
separated by the energy gap equivalent to 170.5 K.   For this source, the observed T$_{10}$ ranges from 90 to 480 K, 
matching with our prediction of the average temperature mentioned earlier. Both the temperatures are cooler than the GRB-DLAs without H$_2$. 
This suggests that the H$_2$-bearing GRB-DLAs trace the cold neutral phase of the ISM. It is to be noted 
that, 
the kinetic temperature is directly measured for H$_2$-bearing DLAs. Whereas, it is not well constrained for non-H$_2$-bearing DLAs.
Though, so far, there is no observational data on 21 cm absorption for this source, we calculate the excitation  temperate of the hyperfine levels 
involved in 21cm transition ($T_{21cm})$ using,  
\begin {equation}
T_{21cm} = -0.068 \left[ln \frac{N_2} {3N_1}\right]^{-1} K.    
\end {equation}
Here, N$_2$ and N$_1$ are column densities of the hyperfine levels $^{1}S_{1/2}$ and $^{0}S_{1/2}$ \citep{{1981Urbaniak}, {1985Deguchi}, {1959Field}}, respectively,  separated by the energy gap equivalent to 0.068 K. Previously \citet{2017Shaw} 
described all the possible excitation and de-excitation processes of the hyperfine levels related to 21 cm transition, as incorporated by CLOUDY. It includes 
collisional excitations and de-excitations by H$^0$, e$^{-}$ and H$^{+}$, together with the non-thermal radiative processes 
including direct pumping by the 21 cm radio continuum, pumping by the continuum near Ly$\alpha$, and by the Ly$\alpha$ line itself.   
Here we find $T_{21cm}$ to be 508 K and 1006 K for case 1 and case 2, respectively. This is higher than the kinetic temperature. 
Similar trend had also been observed for the QSO-DLA at redshift 1.78 along Q1333+ \citep{{2000Chengalur}, {2005Cui}}. 

In our models the gas temperature is determined by heating and cooling balance where various processes contribute to the total heating and cooling.
We find that the grain photoelectric heating is the dominant heating processes for this source. In the grain photoelectric heating, a dust grain 
absorbs a FUV photon and emits 
energetic electrons which heats up its surrounding region by collisions. We find that the grain photoelectric heating 
diminishes deep inside the cloud as FUV decreases inside the cloud. On the other hand, fractional heating by cosmic ray ionisation increases 
as one moves inside the cloud. The dominant coolants are found to be O I and C II lines. 

We would like to emphasis here that for both cases we find the GRB-DLA 120815A is denser, smaller, and cooler than that reported by \citet{2009Ledoux} for GRB-DLAs with no H$_{2}$. Next we discuss GRB-DLA 121024A.



\subsection{GRB-DLA 121024A}
\label{subsec:GRB-DLA 121024A} 
In this section, we present our findings for another H$_2$-bearing GRB-DLA: GRB-DLA 121024A. The first trigger of GRB-DLA 121024A was reported in 2012. 
Later, \citet {2015Friis} observed numerous absorption and emission lines spanned over five components. However, H$_2$ was detected in only three out of 
five components (c+d+e) in rest frame Lyman and Werner band. 
They derived H$_2$ in its lowest four rotational states of the ground vibrational level ($\textit J$ = 0, 1, 2, 3). However, CO (< 14.4) was not detected 
even though H$_2$ was detected. In addition to these, many metal lines of the metals Fe, Ni, Cr, Mn, Ca, Zn were present in their first 
ionised state. The reported metallicity ([Zn/H]) and
dust depletion ([Zn/Fe]) are observed to be $-0.6 \pm 0.2$ and $0.85 \pm 0.04$, respectively. Using these the dust-to-gas ratio 
is calculated to be $\approx$ 
0.2 of the local ISM value.

Though the component wise column densities were reported for metal lines, the component wise H$_2$ and H I column densities were not provided. Hence, we add up 
the three H$_2$ -bearing components and consider them as a single component, and try to decipher the physical conditions of this 
stratified component. They have mentioned that the total H I column density for this stratified component is 10$^{21.6}$ cm$^{-2}$.  
In this calculation we use the metalicity and dust-to-gas 
ratio to the observed value. 
\citet{2015Friis} used -0.9$\pm$0.02 as the spectral index for the GRB afterglow. In addition to this, \citet{2018Li} 
also observed -0.86$\pm$0.01 as the spectral index for GRB 121024A. Keeping the observed value in mind,
we fix the slope for the 
power law continuum to -0.9.
Similar to our study of previous DLA, here also in our calculations we consider 
Case 1 and 2. The individual components of the SED have similar features as shown in Fig. \ref{fig:fig_conti}. The equivalent UV fields are 15 and 28 Habing 
for cases 1 and 2, respectively.
Below we presents our results for these two cases.

We list our best model parameters for these cases in the second and third columns of Table \ref{tab:table 4}.
The abundances of the elements, whose lines have been observed, are varied to match their observed values. 
The predicted hydrogen density is 
100 cm$^{-3}$ and 
60 cm$^{-3}$ for case 1 and case 2, respectively. Beside hydrogen density, the ionisation 
parameter for the powerlaw continuum also differs for these two cases. Our best model for case 1 and case 2 predicts value of A$_v$  $\approx$ 0.65 
and 0.54, respectively. This value is consistent with A$_v$ = 0.9$\pm$0.3 as measured by
    \citet {2015Friis}. Our best models predict the cosmic ray ionisation rate for hydrogen 0.36 $\times$ 10$^{-16}$ s$^{-1}$, same for both the cases. 
    
\begin{figure}
\vspace{-0.3in}
\hspace*{-0.3in}
\includegraphics[scale=0.4]{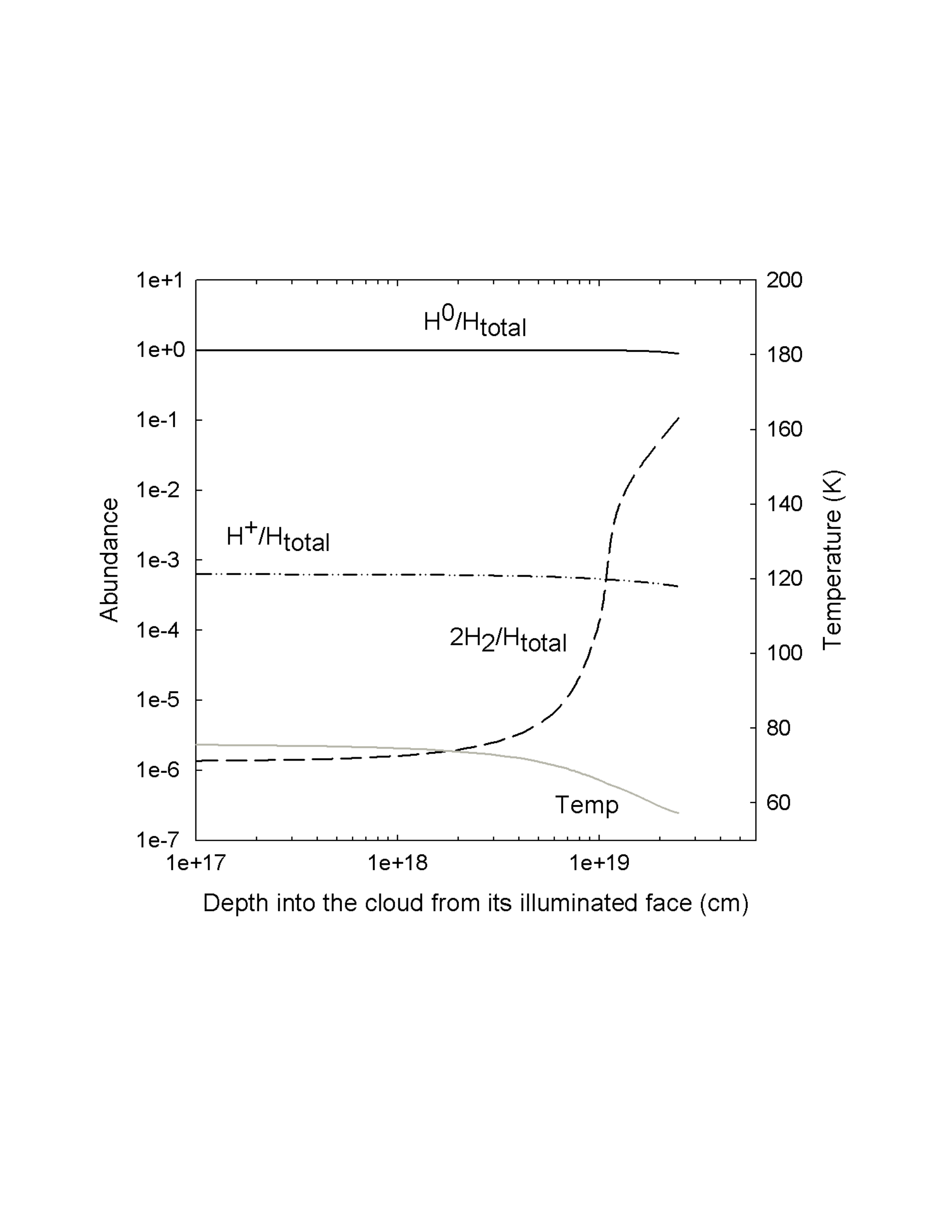}
\vspace*{-0.9in}
    \caption{Temperature and H$^0$/H$_{total}$, H$^+$/H$_{total}$, 2H$_2$/H$_{total}$ are plotted as a function of distance into GRB-DLA 121024A (case 1). The temperature axis
    is shown in the right. This plot shows
    half of the cloud. The other half is just the reflection of this portion.}
    \label{fig:fig_ad2}
\end{figure}

\begin{figure}
\vspace{-1.0in}
 \hspace{-0.3in}
\includegraphics[scale=0.4]{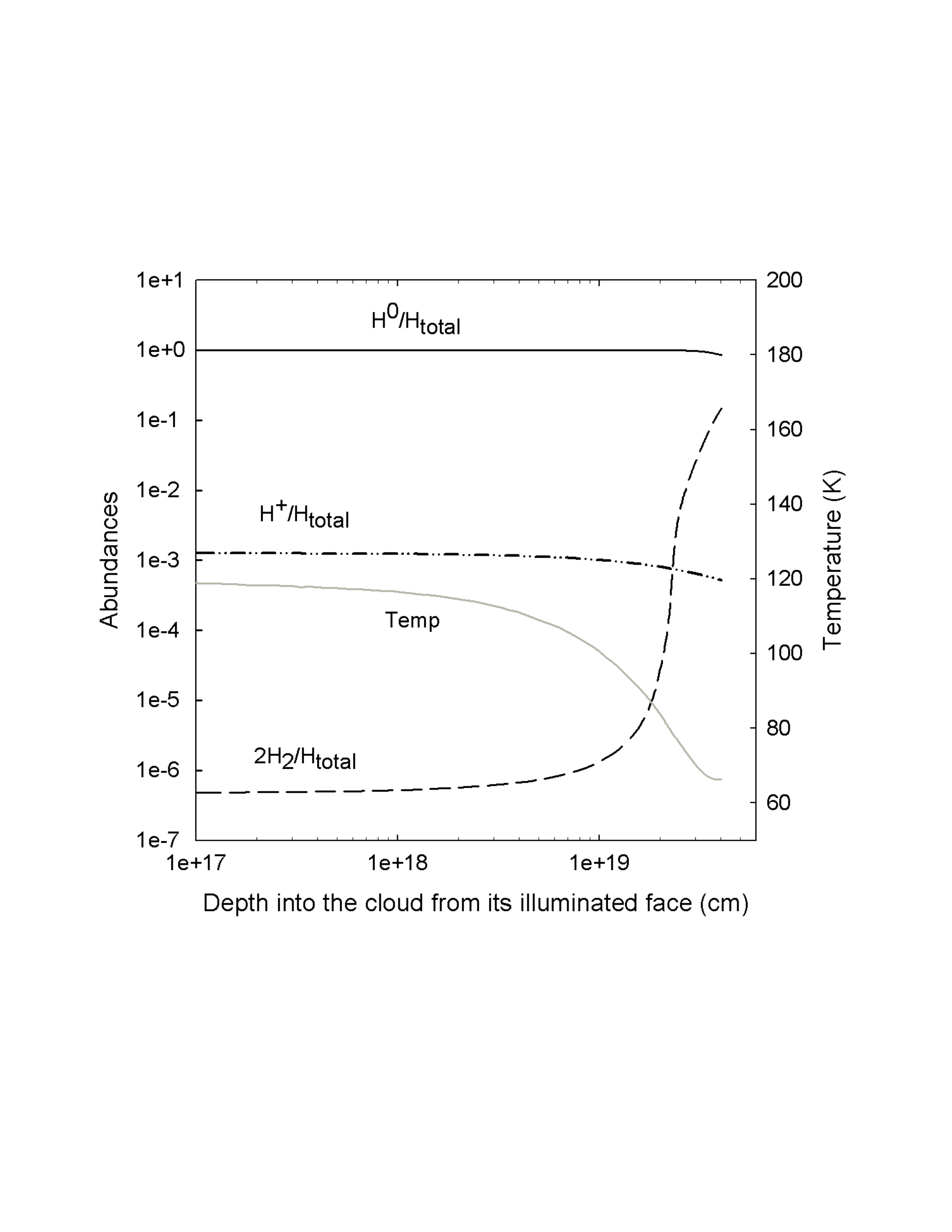}
\vspace*{-1.3in}
    \caption{Temperature and H$^0$/H$_{total}$, H$^+$/H$_{total}$, 2H$_2$/H$_{total}$ are plotted as a function of distance into GRB-DLA 121024A (Case 2). 
    The temperature axis
    is shown in the right.}
    \label{fig:fig_as2}
\end{figure}

In the fourth and fifth columns of  Table \ref{tab:table 5} we compare our predicted column densities with the observed data 
for case 1 and case 2, respectively. 
As can be seen that our predicted column densities match very well with the observed values within the observed error bars.
Previously \citet {2015Friis} reported column densities for two different $b$ parameters, 1 and 10 km s$^{-1}$. 
Their $\textit J$ =2, 3 column 
densities differ by more than a dex for $b = 1$ and 10 km s$^{-1}$. Our best models uses micro-turbulence $b = 1$ km s$^{-1}$ and the predicted 
$\textit J =2, 3$ column 
densities lie between their values. We check a model with $b$ = 10 km s$^{-1}$, but the results for $\textit J =2, 3$ column 
densities do not differ much.
Keeping in mind, the 
uncertainty 
of dust grain sizes at high-redshift, two additional models are also considered with half and double the size of ISM grains. However, none of those
give the desired effect of reproducing the observed  $\textit J =2, 3$ column 
densities. In Table 6 we present the 
change in predicted column densities (in dex) by varying important
parameters by 0.3 dex around the value used in
the best model. Since the result is similar for both the cases, we show results for Case 1 only. It is to be noted that like the previous GRB-DLA, 
among all the parameters, the effect of dust grains is the maximum.  
Temperature (solid grey), and  abundances of 
H$^0$/H$_{total}$ (solid-black) , H$^+$/H$_{total}$ (black-dotted) , 2H$_2$/H$_{total}$ (black-dashed) 
are plotted in Figs. \ref{fig:fig_ad2} and \ref{fig:fig_as2} as a function of distance into the cloud for case 1 and case 2, respectively. 
It can be seen that the H$_2$ fraction starts to increase beyond 
10$^{19}$ cm from the illuminated face of the cloud.
For our best model of case 1, temperature varies between 73 K to 57 K with a mean of 65 K averaged over 
the linear thickness of the cloud, 
which matches well with the observed excitation
temperature T$_{10}$ at 60K. 
We also calculate the excitation
temperature of the hyperfine levels involved in 21 cm transition and found it to be  $T_{21cm} = 1370$ K, which is much higher 
than the kinetic temperature.
Furthermore, our best model predicts the linear size of the GRB-DLA 121024A to be nearly 16 parsec for case 1.

\begin{table}
	\centering
	\caption{Physical parameters of GRB-DLA 121024A using CLOUDY.}
	\label{tab:table 4}
	\begin{tabular}{lcr} 
		\hline
		Physical parameters & best values & best values\\
		&Case 1 &Case 2\\
		\hline
		Power law: log($\chi$) & -3.3 & -2.8\\
		Black body: Temp (log K), log($\chi$) & 4.9, -2.5 & 4.9, -2.5 \\
		Density n(H) ($cm^{-3}$) & 100 & 60\\
		Cosmic ray ionisation rate (10$^{-16}$ s$^{-1}$) & 0.4  & 0.4 \\
		$[$Fe/H$]$ & -1.6 & -1.6\\
		$[$Mg/H$]$ & -1.0 & -1.0\\
		$[$Ca/H$]$ & -2.6 & -2.6\\
		$[$Zn/H$]$ & -0.9 & -0.9\\
		$[$Ni/H$]$ & -1.5 & -1.5\\
		$[$Mn/H$]$ & -1.8 & -1.8\\
		$[$Cr/H$]$ & -1.4 & -1.4 \\
		\hline
	\end{tabular}
\end{table}

\begin{table}
	\centering
	\caption{Comparison of observed and predicted column densities GRB-DLA 121024A (in log scale) using CLOUDY.}
	\label{tab:table 5}
	\begin{tabular}{lcccr} 
		\hline
		Species  & observed & observed  & predicted & predicted\\
		& b=10 & b=1 & b=1 & b=1 \\
		 &(cm$^{-2}$)& (cm$^{-2}$) & (cm$^{-2}$)& (cm$^{-2}$)\\
		 &&&Case1 &Case2\\
		\hline
		H I &--& 21.6 & 21.77 & 21.69\\
		H$_2$ &--& 19.90 & 19.90 & 19.90\\ 
		Mg I &--& <13.86 & 13.56 & 13.44\\
		Ni II &--& 14.47 $\pm$ 0.06 & 14.54 & 14.45 \\
		Zn II &--& 13.47 $\pm$ 0.04 & 13.52 & 13.43\\
		Ca II &--& 12.40 $\pm$ 0.12 & 12.40 & 12.37\\
		Fe II &--& 15.58 $\pm$ 0.03 & 15.63 & 15.55\\
		Mn II &--& 13.47 $\pm$ 0.03 & 13.51 & 13.43\\
		Cr II &--& 13.97 $\pm$ 0.03 & 14.03 & 13.94\\
		CO &--&--& 10.93 & 11.26\\
		HCl &--& --& 12.28 & 12.18\\
		H$_2$(0) &19.7  &19.7 & 19.69 & 19.63\\
		H$_2$(1) & 19.2 &19.3 & 19.47 & 19.55\\
		H$_2$(2) & 16.1 &18.3 & 17.72 & 17.83 \\
		H$_2$(3) & 16.0 & 18.2& 17.21 & 17.83\\
		H$_2$(total) &19.8  &19.9 & 19.90 & 19.90\\
		\hline		
	\end{tabular}
\end{table}

\begin{table}
	\centering
	\caption{Variation in column density of GRB-DLA 121024A (in dex) for variation in different input parameters (Case 1).}
	\label{tab:table 6}
	\begin{tabular}{lccr} 
		\hline
		Species  & $\Delta$n(H) & $\Delta$grain   & $\Delta$log($\chi$)\\
		&&&power law\\
		  0.3 dex& 0.3 dex &0.3 dex & $-$0.3 dex\\
		\hline
		H I      & 0.02 & $-$0.31  &$-$0.15\\
		Mg I     &0.02    & $-$0.37   &$-$0.06  \\
		Ni II    &0 .01    & $-$0.31  &$-$0.15\\
		Zn II    &0.02  & $-$0.30  &$-$0.15 \\
		Ca II    &0.04  & $-$0.35  &$-$0.01 \\
		Fe II    &0.02 & $-$0.33  &$-$0.15 \\
		Mn II    &0.03     & $-$0.29  &$-$0.15 \\
		Cr II    &0.02  & $-$0.30  &$-$0.15 \\
		CO       &0.16  &  0.71  & 0 \\
		HCl      & 0.01&  0.08  & 0.15 \\
		H$_2$(0) &-0.03 & $-$0.16  &0.02\\
		H$_2$(1) &0.04  &  0.18  &$-$0.10 \\
		H$_2$(2) &0.06  &  0.17  &$-$0.15\\
		H$_2$(3) &0.26  &  0.03  &$-$0.29\\
		\hline		
	\end{tabular}
\end{table}

For the 
one-sided radiation field model, the predicted size of the DLA is found to be nearly 13 parsec. Also we find the gas temperature is 111 K at illuminated face of the cloud and it decreases to 
66 K at the shielded face of the cloud with an average of 86 K over the linear size of the cloud. Furthermore we find that like 
the previous source, here $T_{21cm}$ $\approx$ 1800 K, which is higher than the kinetic temperature. It is to be noted 
that here also both the gas temperature and the $T_{21cm}$ is higher for Case 2 than Case 1. 
For both the cases, the grain photoelectric heating is 
found to be the main source of heating, whereas the main sources of cooling points to C II line.   
Like the previous GRB-DLA, this GRB-DLA is also denser, smaller, and cooler than that reported by \citet{2009Ledoux} for GRB-DLAs with no H$_{2}$.

\section{Summary and Conclusions}
GRB afterglows provide an unique opportunity to study the
interstellar medium of star-forming galaxies at high-redshift. Observationally many GRB-DLA show H$_2$, the main constituent of molecular clouds
where star formation takes place.
However, till date there is no detail study, incorporating quantum mechanical microphysics, on these systems to model H$_2$ lines. In this work for the first time we carry out such a study using spectral synthesis code CLOUDY, by modelling
the observed column densities \citep{2015Friis} of GRB-DLA 121024A and 
GRB-DLA 120815 \citep{2013Kruhler}self-consistently. We select these two specific H$_2$-bearing GRB-DLAs, situated at redshifts 2.30 and 2.36 respectively, 
as they have substantially large H$_2$ molecular fraction. While the gas is exposed to UV radiation from everywhere it is exposed to GRB afterglow only from 
one side. Since the current CLOUDY set up does not allow to include
both one-sided and two-sided radiation field simultaneously in a given model,  we consider two cases separately. In Case 1, we assume that the gas cloud is irradiated from both sides, whereas in Case 2 
we assume that the gas cloud is irradiated from one side.
We present results for both the cases. The first number in the parenthesis represents Case 1 and the second number represent Case 2, respectively. 
We will follow this convention throughout.
\begin{itemize}
\item \noindent The total hydrogen density, consisting of ionised and atomic hydrogen together with all the hydrogen bearing molecules, are found to be 
(440, 230) and 
(100, 60) cm$^{-3}$ for GRB-DLAs 120815 and 121024A, respectively.
Earlier, \citet{2009Ledoux} estimated a particle density of 5-15 cm$^{-3}$ for 
seven z > 1.8 GRB afterglows of their sample which did not show H$_2$. Hence, our findings strongly suggest that the total hydrogen densities 
of H$_2$-bearing GRB-DLAs are higher than that of without H$_2$ GRB-DLAs. This finding is also very much consistent with the recent observation of 
\citet{2019Bolmer} who indicated that the GRB-DLA's gas pressure is higher and conversion of H I to H$_2$ occurs there.
\item \noindent Earlier, \citet{2009Ledoux} estimated 
    a linear cloud size of 520$^{+240}_{-190}$ pc for GRB-DLAs lacking H$_{2}$.
    Here we find that linear sizes of H$_2$-bearing GRB-DLAs, 120815A and 121024A, are (9.8, 17.7) and (16, 13) pc, respectively.  
    Taking together this also suggests that the linear sizes of H$_2$-bearing GRB-DLAs are smaller than that without H$_2$ GRB-DLAs. 
    \citet{2018Noterdaeme} had also concluded that H$_{2}$-bearing QSO-DLAs have small physical extent.
 \item  \noindent \citet{2009Ledoux} estimated 
 that the kinetic temperatures of the metal-poor H$_2$ lacking GRB-DLAs are more than 1000 K. Our results suggest that the gas temperatures of H$_2$-bearing GRB-DLAs are lower than that 
 of GRB-DLAs without H$_2$.
\item \noindent The higher density, lower temperature and smaller physical extension as mentioned above suggests that the H$_2$-bearing 
GRB-DLAs trace the cold neutral phase of the ISM.
\item  \noindent Most of the heating to the environment of these systems is contributed by photoelectric heating.  
\item  \noindent Most of the cooling of these systems is contributed by O I and C II lines.
\item  \noindent We find A$_v$ values (0.32, 0.31) and (0.65, 0.54) for GRB-DLAs 120815 and 121024A, respectively. A$_v$ > 0.1 for both the 
sight lines are consistent with the measurement of \citet{2019Bolmer}.
\item  \noindent The cosmic ray ionization rate for hydrogen along the lines of sight for GRB-DLAs 120815 and 121024A are
 (1.9$\times 10^{-16}$, 1.9$\times 10^{-16}$) s$^{-1}$ and (0.4$\times 10^{-16}$, 0.4$\times 10^{-16}$) s$^{-1}$, respectively.
\item \noindent The predicted  column densities match very well with observed data except for the H$_2$ ($\textit J$ ) = 2 level for GRB-DLA 120815A. 
In case of GRB-DLAs 
121024A, Case 1 predicts H$_2$ ($\textit J$ ) = 0, 1 levels better whereas Case 2 predicts H$_2$ ($\textit J$ ) = 2, 3 levels better. Hence,
only the temperature of these sources are constrained by the H$_2$ ($\textit J$ ) = 0, 1 levels.
\item  \noindent Besides the observed species, the best model for GRB-DLA 120815A predicts OH and OH$^+$ column densities to be more than 10$^{13}$ cm$^{-3}$. 
It will be interesting if this prediction gets verified by future observations.
\item  \noindent We find that the 21 cm spin temperature is higher than the gas kinetic temperature for both these sources, though observationally the 
21 cm absorption 
line has not been found yet for none of them. It is to be noted that similar trend had also been observed for the QSO-DLA at redshift 1.78 along 
Q1333+ \citep{{2000Chengalur}, {2005Cui}}. In future, 21 cm absorption can be observed for GRB-DLAs using LOFAR or SKA.
\end{itemize}

In future, following this study, we will investigate more such H$_2$-bearing DLAs, some of which have recently been observed \citep{2019Bolmer}. It will be interesting to find whether our results also hold in general for all other H$_2$-bearing DLAs. That may help us to gain significant insights into the physical conditions of these important astrophysical systems and to understand why they possess molecular hydrogen, the main constituent of
molecular clouds where star formation takes place. That will be an important contribution to the physics of star formation at high-red shift.

\section*{Acknowledgements}
Gargi Shaw acknowledges WOS-A grant (SR/WOS-A/PM-9/2017) from the Department of Science and Technology, India and also like to thank 
the Department of Astronomy and Astrophysics, TIFR for its support. We would also like to 
thank the referee for his/her valuable comments and suggestions.











\bsp	
\label{lastpage}
\end{document}